\documentclass{article}

\usepackage{arxiv}
\usepackage{abbreviations}

\usepackage[utf8]{inputenc} % allow utf-8 input
\usepackage[T1]{fontenc}    % use 8-bit T1 fonts
\usepackage{hyperref}       % hyperlinks
\usepackage{url}            % simple URL typesetting
\usepackage{booktabs}       % professional-quality tables
\usepackage{amsfonts}       % blackboard math symbols
\usepackage{nicefrac}       % compact symbols for 1/2, etc.
\usepackage{microtype}      % microtypography
\usepackage{lipsum}
\usepackage{graphicx}
\usepackage[square,sort,comma,numbers]{natbib}
\usepackage[dvipsnames]{xcolor}
\usepackage{amssymb} % To allow \lesssim symbol

\hypersetup{%
  colorlinks=true,% hyperlinks will be coloured
  linkcolor=blue,% hyperlink text will be blue
  citecolor=blue,
  urlcolor=blue
}

\def \arcsec {$^{\prime\prime}$}
\def \arcmin {$^\prime$}

\def \mum {$\mu$m}

\title{Magnetic Fields studies in the next decade}

\author{
  Ray S. Furuya\thanks{\texttt{rsf@tokushima-u.ac.jp}}$\:\:^{1}$~~$\bullet$~~
  Kate Pattle$^{2}$~~$\bullet$~~
  Simon Coud\'e$^{3}$~~$\bullet$~~ 
  Tao-Chung Ching$^{4}$~~$\bullet$~~
  Steve Mairs$^{5}$\\
  \textbf{Sarah Sadavoy$^{6,7}$~~$\bullet$~~
  Peter Scicluna$^{8}$~~$\bullet$~~
  Archana Soam$^{3}$~~$\bullet$~~
  Chakali Eswaraiah$^{4}$~~$\bullet$~~ 
  Samar Safi-Harb$^{9}$}\\\\
  \textit{$^{1}$Institute of Liberal Arts and Sciences, Tokushima University, Minami Jousanajima-machi}\\\textit{1-1, Tokushima 770-8502, Japan}\\
  \textit{$^{2}$Institute of Astronomy and Department of Physics, National Tsing Hua University,}\\\textit{Hsinchu 30013, Taiwan, R.O.C.}\\
  \textit{$^{3}$SOFIA Science Center, USRA
  NASA Ames Research Center
  Moffett Field, CA 94035, U.S.A.}\\
  \textit{$^{4}$CAS Key Laboratory of FAST, 
  National Astronomical Observatories, Chinese Academy of Sciences,}\\\textit{Datun Road, Chaoyang District, Beijing 100101, People's Republic of China}\\
  \textit{$^{5}$East Asian Observatory (JCMT)
  660 N. A`ohoku Place, Hilo, HI, 96720, U.S.A. }\\
  \textit{$^{6}$Harvard-Smithsonian Center for Astrophysics, 60 Garden Street, Cambridge, MA, 02138, USA}\\
  \textit{$^{7}$Department for Physics, Engineering Physics and Astrophysics, Queen’s University,}\\\textit{Kingston, ON, K7L 3N6, Canada}\\
  \textit{$^{8}$Academia Sinica Institute of Astronomy and Astrophysics, AS/NTU Astronomy-Mathematics Building,}\\\textit{No 1. Sec. 4 Roosevelt Rd, Taipei, Taiwan}\\
  \textit{$^{9}$Dept of Physics and Astronomy, Faculty of Science, University of Manitoba, Winnipeg, MR R3T 2N2 Canada}
}

\begin{document}
\maketitle

\begin{abstract}

\renewcommand{\labelitemi}{\textasteriskcentered}

Magnetic fields are ubiquitous in our Universe, but remain poorly understood in many branches of astrophysics.  A key tool for inferring astrophysical magnetic field properties is dust emission polarimetry.
The James Clerk Maxwell Telescope (JCMT) is planning 
a new 850~\mum\ camera consisting of an array of 7272 paired Microwave Kinetic Inductance Detectors (MKIDs), which will inherently acquire linear polarization information.  The camera will allow wide-area polarization mapping of dust emission
at 14\arcsec-resolution, allowing magnetic field properties to be studied in a wide range of environments, including all stages of the star formation process, Asymptotic Giant Branch stellar envelopes and planetary nebula, external galaxies including starburst galaxies and analogues for the Milky Way, and the environments of
active galactic nuclei (AGN).  Time domain studies of AGN and protostellar polarization variability will also become practicable.  Studies of the polarization properties of the interstellar medium will also allow detailed investigation of dust grain properties and physics.  These investigations would benefit from a potential future upgrade adding 450~\mum\ capability to the camera, which would allow inference of spectral indices for polarized dust emission in a range of environments.  The enhanced mapping speed and polarization capabilities of the new camera will transform the JCMT into a true submillimetre polarization survey instrument, offering the potential to revolutionize our understanding of magnetic fields in the cold Universe.
\end{abstract}

\section{Introduction}
\label{sec:intro}

\renewcommand{\labelitemi}{\textasteriskcentered}

Our Universe is threaded by magnetic fields (also known as $B$-fields), whose presence is deduced from their effects on the astrophysical generation of electromagnetic radiation, or on the propagation of that radiation through the interstellar or intergalactic media (ISM and IGM respectively) \cite[e.g.][]{Han2017}, and so
through observation of polarized astrophysical signal.
Magnetic fields can significantly affect the dynamics of all phases of the ISM, being coupled to the neutral material by Alfv\'{e}nic flux freezing (``frozen in") \cite{Alfven1942}.
These magnetic fields may be primordial \cite{Grasso2001} or generated or amplified by dynamo effects \cite{Han2017}, and can be dissipated by magnetic reconnection \cite{Vishniac1999}. 
In order to address some of the most pressing questions in modern astrophysics and cosmology 
we require knowledge of the structure and strength of magnetic fields in the ISM and IGM, the
physical roles that they play, and the conditions under which they affect gas dynamics.\par

Submillimetre emission polarimetry is a key tool for deducing magnetic field properties in cold ($\lesssim 100$\,K) gas.  Polarized continuum emission arises from non-spherical dust grains aligned with their major axes perpendicular to their local magnetic field \cite{DavisGreenstein1951}: a powerful tracer of plane-of-sky ISM magnetic field direction, as  
dust makes up 1\% of the ISM by mass \cite{Bohlin1978},
and is widely used as a proxy for molecular hydrogen \cite{Hildebrand1983}.
Emission polarimetry is unique in its dynamic range and wide mapping area.  Polarized signal is down to column densities $\sim 10^{20}$\,cm$^{-2}$ in space-based observations \cite{PlanckIntXIX2015}, while submillimetre dust emission remains optically thin at even the highest ISM gas densities \cite{Hildebrand1983}.
Dust polarization fraction ranges from a maximum of $\sim 20$\% in the diffuse ISM \cite{PlanckIntXIX2015} to $\lesssim 1$\% in the densest parts of molecular clouds \cite[e.g.][]{Kwon2018}, and so polarization observations require a sensitivity $\gtrsim 10^{2}$ times better than is needed in unpolarized light.

Methods for quantifying the dynamic importance of magnetic fields inferred from emission polarimetry are well-established.  Plane-of-sky magnetic field strength is inferred using the Davis-Chandrasekhar-Fermi (DCF) method \cite{Davis1951, ChandrasekharFermi1953}, which takes deviations in magnetic field angle to result from Alfv\'{e}nic distortion by
non-thermal motions. 
The dynamic importance of the magnetic field relative to gravity is assessed using the mass-to-flux ratio, the critical value of which indicates a structure too massive to be supported by its internal magnetic field,
while importance relative to non-thermal ISM motions is assessed using the Alfv\'{e}n Mach number, the ratio of gas velocity dispersion to Alfv\'{e}n velocity \cite{Alfven1942}. The dynamic importance of magnetic fields can also be characterised through their morphology \cite[e.g][]{Soler2013}.

The James Clerk Maxwell Telescope (JCMT) is a 15\,m telescope operating in the wavelength range $450-1100$\,\mum, with a resolution of 14\arcsec\ at 850 \mum,
near the summit of Mauna Kea in Hawaii.  The JCMT has for decades been a world leader in submillimetre emission polarimetry,
hosting the UKT Polarimeter \cite{Flett1991}, the SCUPOL polarimeter \cite{Murray1997,Greaves2003} on the SCUBA camera \cite{Holland1999}, and now the POL-2 polarimeter \cite{Bastien2011,Friberg2016} on the SCUBA-2 camera \cite{Holland2013}.  Each of these has measured polarization by inserting a half-wave plate
into a camera's light path, progressing from a sensitivity of $\sim 200$\,mJy beam$^{-1}$ in a single pixel \cite[UKT Polarimeter,][]{Flett1991}, to $\sim 1$\,mJy beam$^{-1}$ over $>5000$ pixels \cite[POL-2,][]{Friberg2016,Ward-Thompson2017}.
The JCMT has made the first detections of magnetic fields in protostellar envelopes with the UKT Polarimeter \cite{Minchin1995,Tamura1995}; in the centre of a starburst galaxy \cite{Greaves2000} and in a starless core \cite{Ward-Thompson2000} with SCUPOL;
and in a photoionized column with POL-2 \cite{Pattle2018}.  The JCMT has made most DCF measurements of magnetic field strength in the ISM to date  \cite{PattleFissel2019}.

\begin{figure*}[t!!]
  \centering
  \includegraphics[width=\textwidth]{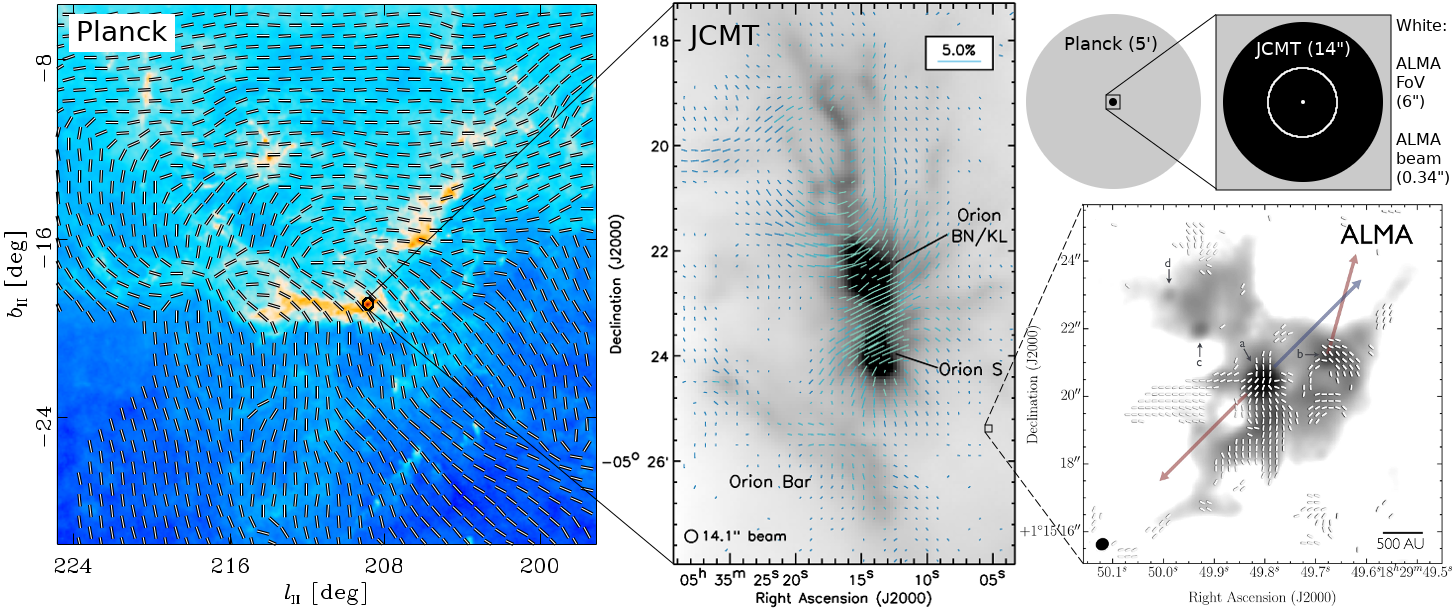}
  \caption{A comparison of JCMT, $Planck$ and ALMA polarization observations.  All vectors are rotated by 90$^{\circ}$ to trace magnetic field direction.  \emph{Left:} $Planck$ observations of the Orion Molecular Cloud \cite{PlanckIntXIX2015}.  \emph{Centre:} JCMT observations of the OMC-1 region in the centre of Orion \cite{Pattle2017}.  Note the significant deviations from the large-scale field morphology.  The total area observed with the JCMT is shown as a black circle on the left-hand panel.  \emph{Lower right:} ALMA observations of the Serpens SMM\,1 protostar \cite{Hull2017},located at comparable distance to Orion
  \cite{Kounkel2017,OrtizLeon2017}.  The extent of the ALMA observations is shown as a black square on the central panel (Serpens MM1 is not located in the marked region).  \emph{Upper right:} The resolutions of $Planck$ (grey), the JCMT (black) and ALMA (filled white circle), and the field of view of ALMA Cycle 7 polarization observations (open white circle; $1/3$ of the 18\arcsec\ ALMA primary beam).  ALMA beam size depends on array configuration; the beam shown is representative of the data in the lower right-hand panel.} 
  \label{fig:scales}
\end{figure*}

Other recent advances have been the $Planck$ Space Observatory \cite{PlanckIntXIX2015} all-sky polarization maps, and the polarimetric capabilities of the Atacama Large Millimeter/submillimeter Array (ALMA) \cite[e.g.][]{Hull17}.
The 5\arcmin-resolution $Planck$ all-sky maps reveal the large-scale polarization structure of the Milky Way, but at best coarsely resolve molecular clouds.  
Conversely, ALMA can map detailed magnetic fields around individual compact objects but,
with a maximum observable size scale $\sim$1\arcsec,
cannot provide larger-scale context.
With a resolution $\sim 10$\arcsec, 
the JCMT bridges this gap (see Figure~\ref{fig:scales}), 
flexibly providing information on how Galactic-scale magnetic fields couple to fields on the smallest scales in the ISM, through both wide-area surveys \cite{Ward-Thompson2017} and high-sensitivity mapping of individual sources \cite[e.g.][]{Yen2019}.

The JCMT is planning a major instrumentation upgrade.  First light for a new 850\mum\ camera is planned for October 2022, with 450\mum\ capability added in 2024.  A new large heterodyne array is planned for 2026. The new camera will have a 12$^{\prime}$ field of view, twice that of SCUBA-2, with a focal plane filled with 3636 pixels, each comprising two Microwave Kinetic Induction Detectors (MKIDs), measuring orthogonal linear polarizations from a single scan observation without a half-wave plate.  Native observation of polarized signal, and the improved capabilities of MKIDs over the SCUBA-2 bolometers, will result in a guaranteed $20\times$ increase in polarization mapping speed over POL-2, and an aspirational $40\times$ increase.  As shown in Figure~\ref{fig:fov}, this will allow entire molecular clouds to be mapped in the time currently required to map a single POL-2 field.  This will transform the JCMT into a true polarimetric survey instrument, while retaining its ability to map sources of particular scientific interest to unprecedented depth in polarized light.

In this white paper we present potential science goals for the new JCMT camera.  While we primarily focus on the 850~$\mu$m polarimetric capabilities of the camera, we also discuss how the proposed studies could be enhanced by 450~$\mu$m polarimetric data.  Where relevant we discuss time-domain magnetic field studies. Section~\ref{sec:850sci} considers star formation and the Galactic ISM; Section~\ref{sec:850sci_agb}, evolved stars and stellar remnants; Section~\ref{sec:850sci_galaxies}, magnetic fields in external galaxies, our own Galactic centre, and active galactic nuclei; Section~\ref{sec:850methods_dust}, dust grain physics; Section~\ref{sec:850methods_lines}, potential synergies with heterodyne instruments; and Section~\ref{sec:instruments}, synergies with other polarimeters.  Section~\ref{sec:summary} summarizes the white paper.

\section{Magnetic fields in star formation}
\label{sec:850sci} 

\textbf{Molecular clouds:}
\label{sec:850sci_clouds}
Stars form in molecular clouds,
the gas dynamics of which are regulated by a, interplay between turbulent pressure, 
magnetic fields and self-gravity \cite[e.g.][]{MacLow2004}.  These cold molecular clouds form out of the warm ISM, which is heated both by supernova shocks and by cosmic rays trapped by the galactic magnetic field \cite{Wolfire1995}.
Proposed formation mechanisms include multiple-shock compression of atomic clouds 
embedded in a weak galactic-scale magnetic field \cite{Inutsuka2015}, or colliding flows in a magnetized warm ISM \cite{Heitsch2009}, among other models.  Thus, galactic-scale magnetic fields may be integral to setting the initial conditions for the formation of stars within molecular clouds.

\begin{figure*}[t]
  \centering
  \includegraphics[width=0.4\textwidth]{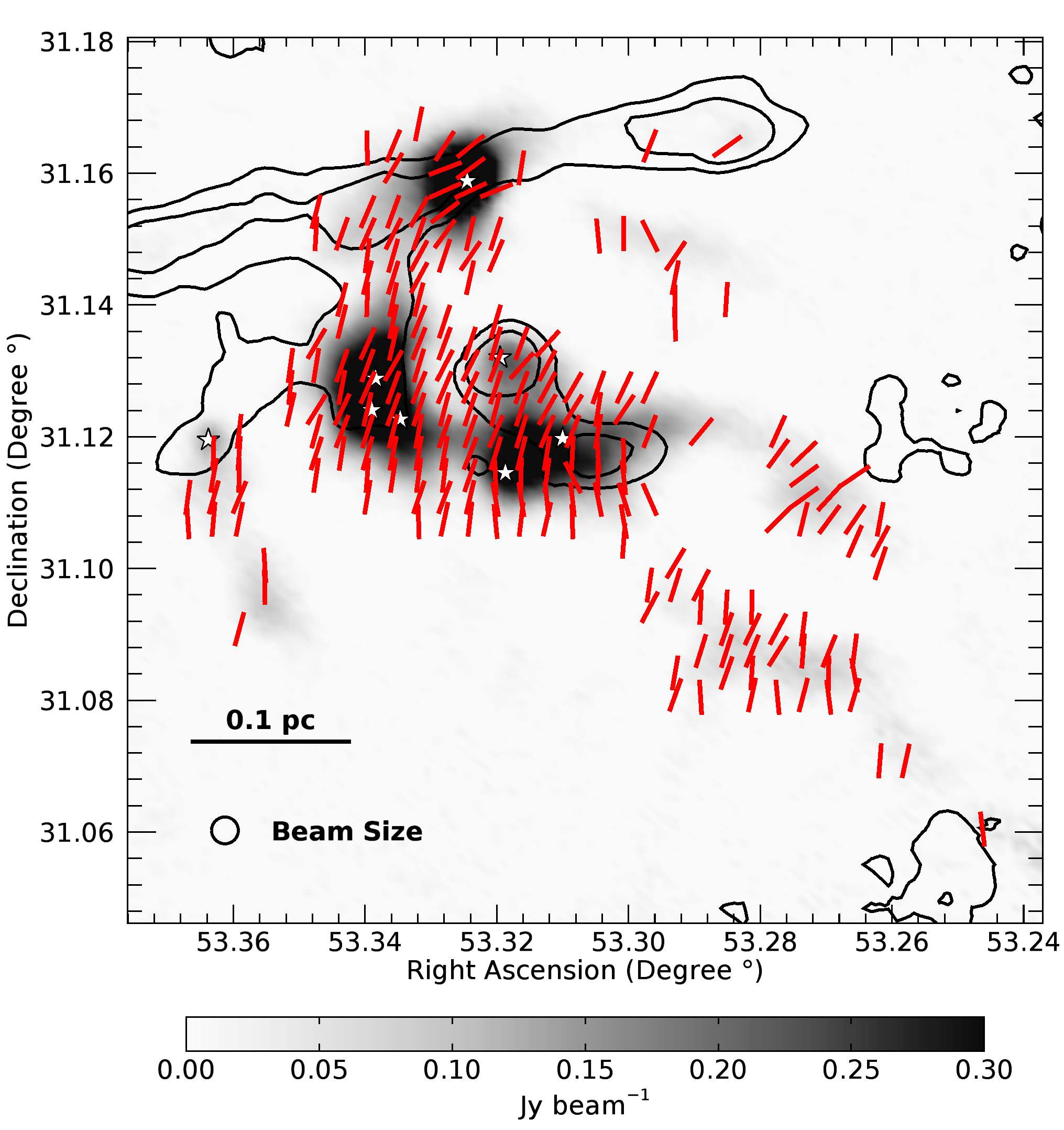}
  \includegraphics[width=0.52\textwidth]{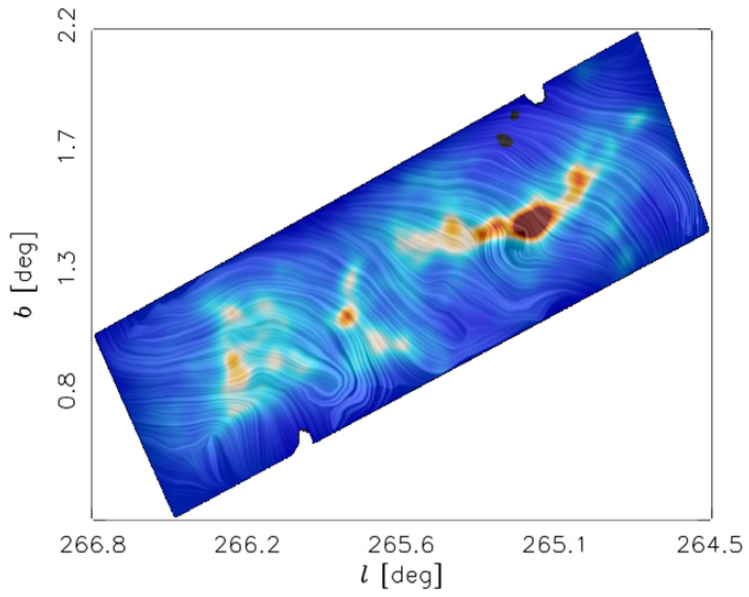}
  \caption{Examples of magnetic fields in molecular clouds. \textit{Left:} Plane-of-sky magnetic field structure observed with POL-2 in the nearby ($\sim 300$~pc) Perseus~B1 low-mass star-forming region \citep{Coude2019}. Image shows 850~$\mu$m intensity; contours show $^{12}$CO~J=3-2 integrated intensity (10 and 20 K\,km\,s$^{-1}$) measured with HARP \citep{Sadavoy2013}. Embedded YSOs are marked with star symbols. \textit{Right:} Plane-of-sky magnetic field morphology observed with BLAST-Pol in the early-stage molecular cloud Vela~C ($\sim 700$~pc) \citep{Fissel2016}.}
  \label{fig:clouds}
\end{figure*}

Both $Planck$ \cite{PlanckIntXIX2015,PlanckIntXXXV2016} and extinction polarimetric observations \cite{Li2013} suggest that large-scale magnetic fields in molecular clouds are bi-modal, being preferentially aligned either parallel or perpendicular to the major axis of the cloud.
The relationship between magnetic fields and filamentary structure within clouds remains uncertain.
Recent observations of the Vela C complex by BLAST-Pol \citep{Jow2018} and of IRDCs by POL-2 \citep{Liu2018}, along with
optical and NIR extinction polarimetric results \cite[e.g.][]{Wang2017, Palmeirim2013}, 
show that magnetic fields on the peripheries of self-gravitating filaments are generally perpendicular to the filaments' major axes (see Figure~\ref{fig:clouds}).  However, these magnetic fields may be aligned with low-density substructures (sometimes called `striations') which are themselves perpendicular to the filaments' major axes \cite{Palmeirim2013}.
These striations may comprise material being accreted onto filaments along magnetic field lines \cite{Andre2014}.  These observations tell us about overall field-filament alignment within molecular clouds, but do not provide sufficient resolution to determine the the behaviour of magnetic field inside filaments.
Conservation of magnetic flux requires that magnetic fields either pass through filaments \cite[e.g.][]{Tomisaka2014} or wrap around them \cite[ e.g.][]{FiegePudritz2000}.
Sensitive high-resolution polarization observations would distinguish between these cases, informing the role of magnetic fields in filamentary accretion and fragmentation.

\begin{figure*}[t]
  \centering
  \includegraphics[width=0.7\textwidth]{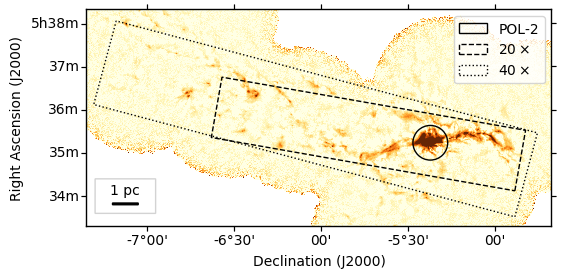}
  \caption{A comparison of polarization mapping speed between POL-2 and the proposed new camera.
  Image shows a SCUBA-2 850~\mum\ map of the Orion A molecular cloud \cite{Salji2015}.
  The `integral filament' is on the right of the image (North is located to image right).  Solid circle shows the full extent of a current POL-2 field, centred on the OMC-1 region.  Rectangles mark the area observable in the same time with $20\times$ (dashed) and $40\times$ (dotted) mapping speed.}
  \label{fig:fov}
\end{figure*}

POL-2 observations of nearby molecular clouds suggest that within dense filaments, relationships between field and filament direction can become more complex.  For example, the centre of the OMC-1 molecular cloud -- the nearest region of high-mass star formation -- shows a field geometry that may have been significantly distorted by large-scale motion of material under gravity \citep[][see Figure~\ref{fig:scales}]{Pattle2017}.  However, the surface-brightness limitation and relatively small extent of a POL-2 observation \cite{Friberg2016} and the insensitivity of SCUBA-2 to large-scale structure \cite{Sadavoy2013} limits our current ability to deduce the properties of magnetic fields within filamentary structure, particularly in low-density non-self-gravitating filaments.  To determine the existence or otherwise of magnetically-supported filaments, higher-sensitivity polarization observations with $< 0.1$\,pc linear resolution are needed \cite{Arzoumanian2011}.  The new 850-$\mu$m camera will allow entire molecular clouds to be mapped in polarized light at 14\arcsec\ resolution, equivalent to $\sim 0.01$\,pc in nearby molecular clouds.

\begin{figure*}[t]
  \centering
  \includegraphics[width=0.32\textwidth]{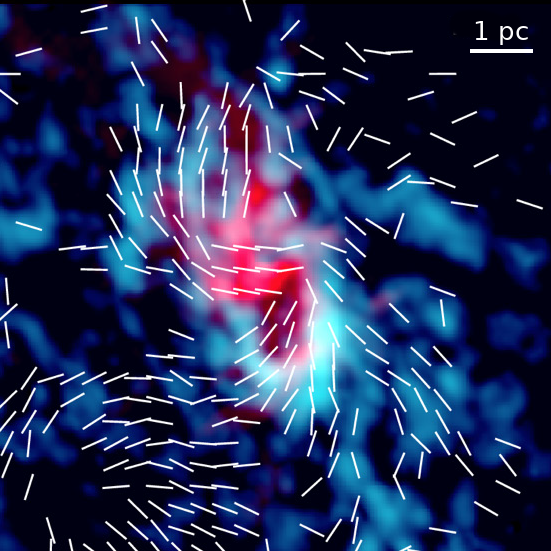}
  \includegraphics[width=0.32\textwidth]{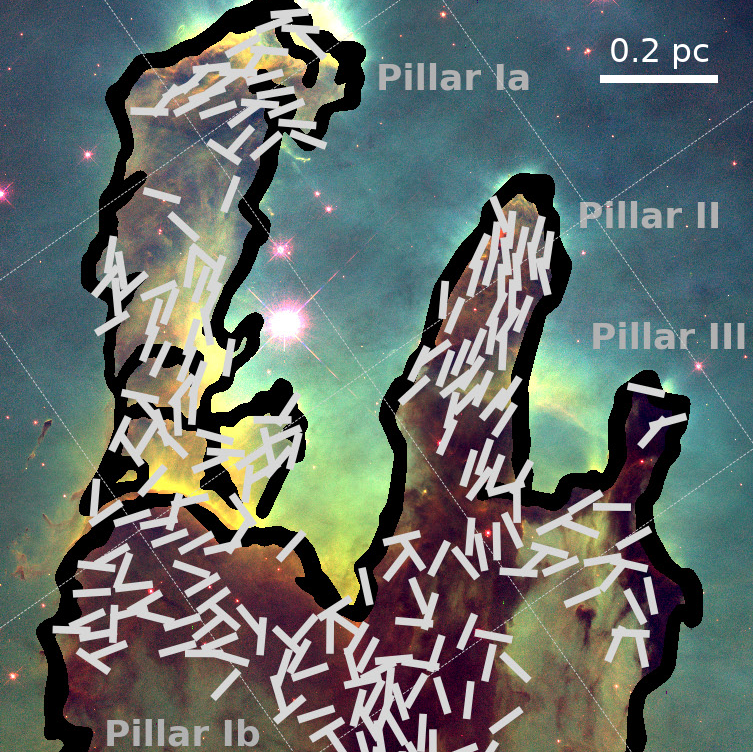}
  \includegraphics[width=0.32\textwidth]{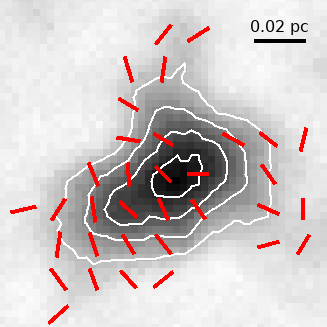}
  \caption{JCMT POL-2 observations of magnetic fields in the Milky Way Galaxy, on a range of scales.  In all panels, POL-2 850\mum\ polarization vectors have uniform length and are rotated to trace magnetic field direction.  \textit{Left}: The centre of the Milky Way, Sagittarius A*, and its circumnuclear disk 
  (SMA imaging; blue) and mini-spiral 
  (6\,cm continuum Jansky Very Large Array imaging; red) \citep{EAO_Newsletter}. \textit{Middle}: The `Pillars of Creation' photoionized columns in M16 (Hubble Space Telescope image) \citep{Pattle2018}. \textit{Right}: The Ophiuchus C starless core (SCUBA-2 850\mum\ image) \citep[c.f.][]{Liu2019}.  }
  \label{fig:cores}
\end{figure*}

Feedback from massive stars drives the dynamics and regulates the evolution of the molecular clouds in which they form \cite{Tan2014}.  
Intense UV radiation and/or winds from OB stars as well as supernova feedback 
drive the expansion of H\textsc{ii} regions, creating structures such as photoionized columns (also known as pillars or elephant trunks) in the photodissociation regions (PDRs) at the interface between molecular and ionized material.  The role of magnetic fields in PDR evolution remains poorly constrained \cite[e.g.][]{Henney2009,MackeyLim2011}.  POL-2 has recently made the first map of magnetic fields within the famous `Pillars of Creation' in M16 \cite[][see Figure~\ref{fig:cores}]{Pattle2018}, finding that the magnetic field is dynamically important,
but unable to prevent the columns' destruction by the oncoming ionization front.  However, the improved mapping speed of the new camera will allow the entire PDR associated with an open H\textsc{ii} region to be mapped in the time currently required to map individual columns, allowing investigation of the role of magnetic fields in the large-scale evolution of H\textsc{ii} regions.

\textit{Time estimate}: Given a 20$\times$ increase in mapping speed, with 14 hours of observing time in Band 2 weather, 0.6 square degrees of the sky could be observed to a depth of 1.5 mJy/beam, as shown in Figure~\ref{fig:fov}. This would allow mapping of full molecular clouds to the depth currently achieved in single pointing observations by the BISTRO (B-Fields in Star-forming Region Observations) Survey \cite{Ward-Thompson2017}, fulfilling the science goals described above.

\textbf{Starless and prestellar cores:}
A key indicator of the relative importance of magnetic fields in the gravitational collapse of cores to form YSOs, and of the magnetic fields in YSOs themselves, is the strength and morphology of magnetic fields in starless cores.  Starless cores are overdensities in star-forming regions which, if gravitationally bound (a `prestellar core' \cite{Ward-Thompson1994}), will go on to form an individual star or system of stars \citep{Benson1989}.  A detailed understanding of how starless cores form and evolve is necessary in order to understand the functional form of the Initial Mass Function \cite{Motte1998}.

Being extended, low-surface-brightness objects, starless cores remain particularly challenging to observe.  The JCMT has been responsible for nearly all polarimetric observations of starless cores to date, both with SCUPOL \citep{Matthews2009} and more recently with POL-2 \citep[e.g.][]{Liu2019}.  Isolated starless cores generally appear to have a smooth and well-ordered magnetic field, with detectable polarization across the cores \cite[e.g.][]{Crutcher2004}.  An example of a starless core observed with POL-2 is shown in Figure~\ref{fig:cores}.  Despite being gravitationally unstable \cite[e.g.][]{Kirk2006}, none of the prestellar cores so far observed unambiguously show the `hourglass' magnetic field which would indicate ambipolar-diffusion-driven collapse \cite{Mouschovias1976}.  The role of magnetic fields in the physics of prestellar core formation and collapse thus remains unclear.  However, very few starless cores have been observed in polarized light, due to the prohibitive amount of time required for a detection, and observations are strongly biased towards the very highest-surface-brightness cores.  With $20\times$ increased mapping speed, the new JCMT camera would allow at least an order of magnitude increase in the number of starless cores detectable, and would allow investigation of whether the uniform fields seen in bright cores are the norm, and to systematically search for cores showing signs of magnetically regulated collapse.

Debate continues over whether high-mass stars form from the monolithic collapse of prestellar cores, analogously to low-mass stars, or through competitive accretion or other dynamic processes \cite{Tan2014}.  If high-mass prestellar cores exist, they are likely to require significant magnetic support \cite[e.g.][]{Pillai2015}.
Archival data from the SCUPOL Legacy Catalogue toward a set of bright massive cores in the G\,11.11-0.12 region
has been used to propose that one of these sources is a magnetically supported high-mass starless core \cite{Pillai2015}. 
Most detections of high-mass star-forming cores to date have been made using interferometric observations
\cite[e.g.][]{Tang2013,Ching2017}
in which any extended lower-density periphery will be resolved out.  The new JCMT camera will allow entire IRDCs to be surveyed in polarized light, searching for polarization geometries consistent with magnetic support, and for existing high-mass core candidates to be surveyed systematically, allowing the debate over magnetically-supported high-mass starless cores to be put onto a statistical footing.

\textit{Time estimate:} Given a 20$\times$ increase in mapping speed, the well-studied prestellar core L1544 (peak 850\mum\ brightness $\sim 300$\,mJy/beam) could be observed to a sensitivity of 0.5 mJy/beam with 5 hours of Band 1 observations, allowing 3-$\sigma$ detection of 0.5\% polarization on-peak. A survey of 20 such cores would thus require only 100 hours of observing time.

\textbf{Protostellar systems:}
Protostellar cores, dense cores with a size $\sim 0.1$\,pc containing embedded hydrostatic objects, either young stellar objects (YSOs) or their precursors, are generally warmer, brighter and more centrally condensed than their starless counterparts, and so are less challenging to observe.
Protostellar cores, containing complex internal structures (discs, accretion flows, etc.), are good interferometric targets \cite{Hull2019}.  However, single-dish observations provide information on the environments of these cores unobtainable with interferometers.  The new JCMT camera offers the opportunity to perform unbiased surveys of the magnetic environments of protostellar cores in nearby molecular clouds.

Recent interferometric observations suggest that the dynamic importance of magnetic fields in protostellar cores may vary widely: the majority have outflows randomly oriented with respect to the magnetic field direction (on scales $\sim$10$^2$ -- 10$^3$ AU), suggesting a weak magnetic field, while a minority show parallel outflow and field directions, suggesting a dynamically important field \cite{Hull2019}.  A large dust polarization survey could investigate whether this behaviour persists on core-to-filament scales ($\gtrsim$0.1\,pc), and whether there is a difference in large-scale magnetic environment between the two populations of protostellar cores.
Such a survey would offer a strong legacy set of data, and would identify targets for interferometric follow-up.
Studying magnetic fields from core scales down to the scales of disks ($\lesssim 10^2$ AU) is vital to understand the origin of those disks and the formation of jets and outflows, in order to determine whether field misalignment, turbulence, or non-ideal magnetohydrodynamic (MHD) processes are at play \citep[e.g.,][]{HennebelleFromang08, Seifried13, Masson16}.

A further unanswered question is of the role played by magnetic fields in the formation of brown dwarfs \cite{Chabrier2014}.  The improved sensitivity and mapping speed of the new JCMT camera will allow for systematic investigation of cloud cores hosting very low luminosity objects (VeLLOs; integrated luminosity $\lesssim 0.1 \, L_{\odot}$),
potential progenitors of either proto-brown dwarfs or very low-mass YSOs \cite[e.g. ][]{Liu2016}. A large sample of VeLLOs could include first hydrostatic cores (FHSCs) -- the much-searched-for adiabatic first kernel of mass which precedes a core's collapse to make a YSO. Regardless of whether VeLLOs are FHSCs or very young YSOs, their outflows are too weak to affect the magnetic fields in their host cores, and so they offer the possibility of mapping the initial field structure in protostellar cores \cite{Soam2015}. 

\textbf{Time-domain science:} A long-standing question in star formation is of the rate at which a YSO gains mass from its surroundings \citep{Hartmann2016}, and of the role of magnetic fields in regulating this process.
While YSOs are inherently variable objects  \citep[e.g.,][]{Evans2009,forbrich2017},
sporadic episodes of elevated mass accretion can be observed at the earliest stages of a YSO's life at far-infrared (FIR) and submillimetre wavelengths \citep{Johnstone2013}. The frequency and amplitude of this variability give insight into the physical drivers of unsteady accretion.  The JCMT Transients Survey first showed that submillimetre protostellar variability is robustly observable \citep{Herczeg2017,Mairs2017b,Johnstone2018}. When a YSO enters a burst phase, the surrounding material reprocesses the excess energy and the submillimetre flux increases with dust temperature \citep{Johnstone2013, Johnstone2018}. Monitoring potential changes in the magnetic field in the accreting material over timescales of weeks to years will give insight into the physical conditions of these systems.
Radio observations have also shown short-timescale (hours) synchrotron flares associated with T~Tauri stars \citep[e.g.][]{Furuya2003, Forbrich2008}. The JCMT recently made the first submillimetre observation of a similar event in the JW~566 T~Tauri Binary System \citep{Mairs2019}, 
thought to be the most powerful of its kind recorded. Fast-followup target-of-opportunity polarimetric observations of the dusty regions associated with such flares will be compared with archival data to note any significant changes in magnetic field properties, even over short timescales.

\textit{Time estimate:} For a YSO with peak 850\mum\ brightness $\sim 1000$\,mJy/beam (such as the variable source EC53 \cite{Yoo2017}), a sensitivity $\sim 3.3$\,mJy/beam would be required in order to make a 3-$\sigma$ detection of 1\% changes in polarization.  Given a 20$\times$ increase in mapping speed, this could be achieved in approximately 10 minutes of Band 2 observing time. This would make both wide-area YSO polarization surveys covering entire molecular clouds and long-term and target-of-opportunity monitoring of protostellar variability feasible.

\textbf{450\,$\mu$m science:} Comparison of 450\,$\mu$m and 850\,$\mu$m measurements of polarization in molecular clouds and cores will allow study of differing polarization structures in warm and cold dust populations along the line of sight, thereby producing quasi-three-dimensional magnetic field models \cite[e.g][]{Seifried13}.
The potential of such studies is demonstrated by recent FIR polarimetric observations of Orion~A,
showing that the magnetic field structure observed at 850\,\mum\ (Figure~\ref{fig:scales}) gives way in the FIR to a polarization structure that traces the bipolar structure produced by the BN/KL explosion \citep{Chuss2019}. 
The 450\mum\ capabilities of the new camera will allow such comparisons to be made as a matter of course.

\section{Late-stage Stellar Evolution}
\label{sec:850sci_agb}

\textbf{AGB stars and planetary nebulae:} The new JCMT camera will significantly improve our knowledge of magnetic fields
in asymptotic giant branch (AGB) stars and planetary nebulae by allowing the study of magnetic fields in circumstellar material, and so of 
the typical magnetic field geometry within a circumstellar envelope,
how magnetic fields regulate mass-loss phenomena in AGB stars (or vice versa), and of
the relationship between stellar and circumstellar magnetic fields.
Cool evolved stars have significant magnetic fields both at their surfaces \citep{Lebre2014} and in their envelopes \citep[e.g.][]{Herpin2006, Vlemmings2011}, measurements of which currently use Zeeman splitting either of atomic lines from the stellar photosphere or of maser transitions of molecules in the circumstellar material. 
However, these measurements sample only a small fraction of the gas associated with the stars,
while photospheric lines may be affected by starspots,
and masers inherently sample only high-density, population-inverted, molecular gas in the inner outflow.
An overall view of magnetism in evolved stars requires observations of the 
overall magnetic field structure of the circumstellar envelope.  

Debate continues as to whether magnetic fields in AGB stars are produced by angular-momentum transfer from a companion.
Statistics of the presence and morphology of the large-scale field will allow comparison to models of the expected population of companions, and with known binary stars.  Comparison between field morphology and  mass-loss history will reveal whether magnetic fields play any role in shaping the outflow.  Correlating magnetic field properties with evolutionary stage will explore how fields evolve with the stars, and if they play a role in the changes that occur as stars evolve off the AGB.  Moreover, along with supernovae, AGB stars have been considered as major sites of dust grain production.  Data provided by the new camera would thus provide new observational constraints on dust grain physics.  

\textit{Time estimate:}
ALMA observations of IRC+10216, the brightest AGB star in the submillimetre \citep{Dharmawardena2018}, suggest that 
POL-2 might need $\sim$ 30 hours of Band 1 weather to detect 5\%-polarized emission in the outer envelope.  A 20$\times$ increase in speed would make the brightest sources observable in 1--2 hours each.
A large program could thus observe tens or hundreds of evolved stars and map the polarization in their envelopes, particularly if informed by the results of the ongoing Nearby Evolved Stars Survey (NESS) Large Program, or by future continuum mapping with the new camera.

\textbf{Supernova remnants:} ISM properties control galactic evolution by regulating star formation rate, while stars return much of their material to the ISM through dense winds or supernova explosions at the end of their lives. These supernovae send shock waves into the ISM, producing supernova remnants (SNRs) which disperse heavy elements, while also compressing and seeding magnetic field lines \citep[e.g.,][]{Leao2009, West2016}. 
The origin of magnetic fields in SNRs and their link to the magnetic field of their host galaxy is an important open question,
with few objects studied in detail \citep[e.g.,][]{Rho2018, WT2017}.
While submillimetre observations offer a new, independent way of probing magnetic fields, only a few SNRs have been mapped in submillimetre polarization to date.With the 20$\times$ increase in mapping speed of the new JCMT camera, 850\,$\mu$m polarization observations of SNRs will be achievable for large numbers of objects, opening a new field of study for the JCMT.  A polarization survey of the nearby galaxies M31 and M33 will help link our understanding of the global view of their star formation with their magnetic field morphology: while most extragalactic SNRs will be very compact, their high polarization fractions mean they should be detectable with the new JCMT camera.

The new camera will also enable novel studies in the field of pulsar wind nebulae (PWNe), a subclass of core-collapse supernovae. These are non-thermal, polarized synchrotron bubbles inflated by the loss of rotational energy from fast-spinning neutron stars.
It is believed that dust grains are able to penetrate into the nebula given the low pulsar velocity, thus making circum-pulsar disks \citep{Lohmer2004}. Future observations of a large sample of PWNe (in their different stages of evolution) will open a new window into the discovery of circum-pulsar disks in which planets may form.

SNR magnetic field studies are important not only to understand the particle acceleration mechanism operating in SNR shocks, but also to address the larger questions of cosmic magnetism and the origin of cosmic rays driving future large radio telescopes such as the Square Kilometer Array and the next generation VLA (ngVLA), and the $\gamma$-ray Cherenkov Telescope Array.  Observations made with the new JCMT camera will serve as pathfinder science for these instruments.

\textit{Time estimate:} 850\mum\ polarization is detected in 9 hours of mixed Band 1/2 POL-2 commissioning observations of the Crab Nebula SNR.  A similar detection would thus be achievable in approximately 30 minutes with a $20\times$ increase in mapping speed.  Fainter SNRs would thus be detectable with a few hours of observing time.

\section{Galactic-scale magnetic fields}
\label{sec:850sci_galaxies}

\textbf{Spiral Galaxies:} The disk of our galaxy is threaded by a large-scale magnetic field, mostly parallel to its spiral arms \citep[e.g.,][]{Fosalba2002, PlanckIntXIX2015}. Similar fields have also been observed in nearby galaxies 
\citep{Fletcher2011,Beck2015a}, indicating that these fields are closely tied to the dynamics of spiral galaxies \citep{Beck2015b}, and are likely sustained by a dynamo effect created by differential rotation and star formation occurring within them \citep[e.g.,][]{Beck1996}. However, Faraday rotation measures
have shown the existence of a field reversal in the inner region of our galaxy \citep[e.g.,][]{Ordog2017}, which has not yet been observed elsewhere, and which could indicate anisotropic turbulence in galactic magnetic fields, or perturbation by satellite galaxies \citep{Beck2015b}. 

Submillimetre dust polarization observations in nearby spiral galaxies will allow insight into the magnetic properties of the high-density gas, which will serve as templates to better understand our own galaxy's magnetic field \citep{PlanckIntXIX2015}. Several of these nearby galaxies were successfully detected at 450~$\mu$m and 850~$\mu$m by the JINGLE survey using the SCUBA-2 camera on the JCMT \citep{Saintonge2018}, and so the new camera will have the required sensitivity to detect polarization in these objects.  The new camera's larger field of view will allow the first 850~$\mu$m polarization surveys of the M31 and M33 galaxies, as discussed above. These extragalactic polarization data sets can be analyzed similarly to those of molecular clouds, providing information about the magnetic and turbulent properties of galactic-scale magnetic fields \cite{Houde2013}.

\textit{Time estimate:} M31 occupies approximately 3 square degrees on the sky.  To map this area to 1.5\,mJy/beam sensitivity in 850\mum\ polarized light with the new camera would require approximately 70 hours of Band 2 time, making polarization surveys of nearby spiral galaxies eminently feasible.
 
\textbf{Starburst Galaxies:} Galactic magnetic field strengths $\sim$100\,$\mu$G are observed in starburst (intensely star-forming) galaxies \citep[e.g.,][]{Adebahr2013}. In comparison, the Milky Way's large-scale field strength is $\sim 5$~$\mu$G, similar to M31 and M33 \citep{Beck2015b}. While the origin of starburst galaxies' strong magnetic fields is not well-understood, their interaction with galactic outflows may have magnetized the IGM in the early universe \citep[e.g.,][]{Bertone2006}. The new JCMT camera will significantly expand our knowledge of the magnetic field structure in the densest regions of starburst galaxies, helping to explain the nature of these fields and how they are maintained in environments of intense stellar feedback \citep[e.g.,][]{Greaves2000,Beck2015b, Jones2019}.

\textit{Time estimate:} The starburst galaxy M82 has a peak 850\mum\ brightness of 1400 mJy/beam and a median polarization fraction of 2.8\% \cite{Greaves2000}.  A 5-$\sigma$ detection of this polarization fraction could be achieved in less than 3 minutes in Band 2 weather with the new camera, and fainter starbursts could be observed in minutes or hours.

\textbf{Super-Massive Black Holes and Active Galactic Nuclei:} The role of magnetic fields in galactic evolution can be investigated through observations of Sagittarius~A*, the super-massive black hole (SMBH) at the centre of our galaxy.  While Sgr~A* is currently quiescent, its accretion behaviour provides information on AGN physics, such as jet launching mechanisms and galactic-scale feedback, unobtainable in more distant sources. Similarly to AGN such as Cygnus~A \citep{Lopez2018Cygnus}, Sgr~A* hosts a circumnuclear disc (CND) with a rotating molecular torus housing ionized streamers \citep{Vincent2015, Vincent2019}.  POL-2 observations of the CND (Figure~\ref{fig:cores}) show that the magnetic field and the CND align on larger scales, while the innermost field lines align with the streamers \cite{EAO_Newsletter,Hsieh2018}, suggesting that the CND and the streamers are an inflow system.  Observations with the new JCMT camera will allow the large-scale magnetic environment of the Galactic centre and Sgr~A* to be mapped in unprecedented detail. These combined with observations of Sgr~A* itself from the Event Horizon Telescope \cite{EHT,Issaoun2019}, of which the JCMT is a part, will revolutionize our understanding of how SMBHs acquire mass, and so how magnetic fields influence the galactic-scale feedback effects that regulate galactic evolution.

\textit{Time estimate:} The POL-2 polarization vectors of the Galactic Centre shown in Figure~\ref{fig:scales} were achieved in $\sim 14.5$ hours of mixed Band 1/2 time, with a sensitivity $\sim 1.6$\,mJy/beam \cite{EAO_Newsletter}.  With the new camera, such an observation could be made in $\sim 45$ minutes, making a deep, wide-area survey of the Galactic Centre quickly practicable.

\textbf{Time-domain science:} Among the most extreme environments in which magnetic fields have been detected are jets launched from accretion events onto SMBHs in radio-loud AGN \citep[e.g.][]{Marscher2006}. The 850~$\mu$m emission of these objects is dominated by highly-polarized synchrotron radiation from relativistic electrons accelerated along magnetic field lines \citep[e.g.][]{Robson2001}, and the turbulent nature of the magnetized medium found within shocks along these relativistic jets may cause their observed temporal variability in polarized intensity \citep[e.g.][]{Jorstad2007, Marscher2014}.  

\textit{Time estimate:} Variable AGN polarization is detected between 40-minute Band 2 850\,\mum\ POL-2 observations (sensitivity $\sim 7$\,mJy/beam) \cite{Friberg2016}. The new camera would decrease the observing time per measurement to 2 minutes, making a daily observing campaign possible.
Such high-cadence measurements would provide a statistically-significant AGN variability data set on timescales of days, and would allow precise measurements of intra-day variability \citep[e.g.][]{Lee2016day}.

\textbf{450\,$\mu$m science:} The smaller beam size of the JCMT at 450~$\mu$m may help to detect polarized emission from radio-loud AGN by reducing the effect of beam dilution on the measured signal. More importantly,
while synchrotron emission is typically the main source of emission in these AGN, their 850~$\mu$m polarization can be contaminated by dust emission,
\citep{Lopez2018}. This dust component would typically be an order of magnitude brighter than the synchrotron emission at 450~$\mu$m, thus lifting the degeneracy between the thermal and non-thermal components at 850~$\mu$m. Combined 450~$\mu$m and 850~$\mu$m polarimetric data of flat-spectrum radio-loud AGN such as blazars can also be used to probe the electron density and magnetic field properties in the inner components of relativistic jets launched by SMBHs \citep[e.g.][]{Jorstad2007,Lee2015,Lee2016day}.

\begin{figure*}[t]
    \centering
    \includegraphics[width=\textwidth]{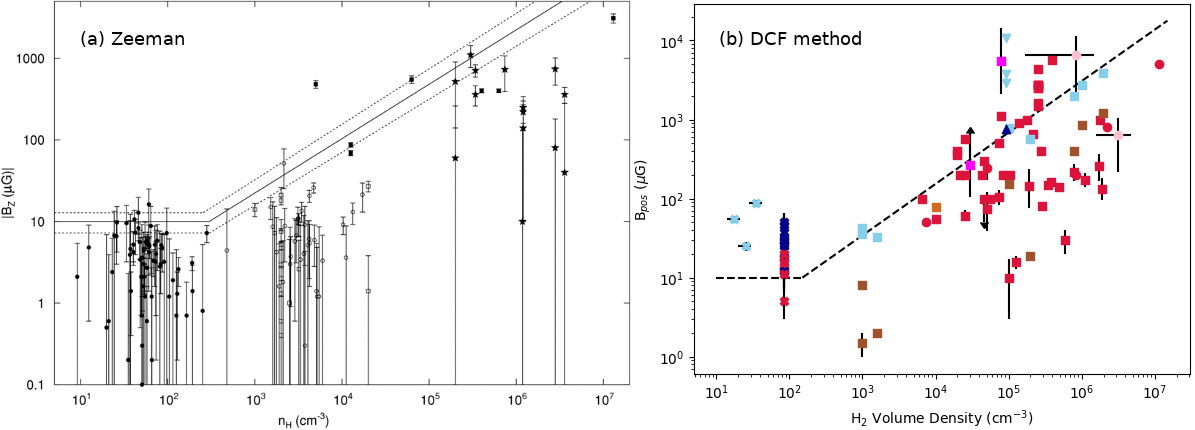}
    \caption{A comparison between magnetic field strengths determined (a) directly from Zeeman splitting measurements \cite{Crutcher2010} and (b) indirectly using the DCF method \cite{PattleFissel2019}.  The left-hand plot compares $H$ volume density to line-of-sight field strength. The right-hand plot compares $H_{2}$ volume density to plane-of-sky field strength.  The dashed line shown in both panels is the upper-limit field strength inferred from the Zeeman measurements in panel (a).}
    \label{fig:zeeman_vs_cf}
\end{figure*}

\section{Dust grain physics and alignment mechanisms}
\label{sec:850methods_dust}

For polarization observations to trace magnetic fields, a fraction of the ISM dust population must consist of non-spherical grains with major axes preferentially aligned perpendicular to the local magnetic field direction \cite{DavisGreenstein1951}.  The most promising theory for how this occurs is the Radiative Alignment Torques (RATs) paradigm \cite{Lazarian2007}, in which irregular grains
are spun up by anisotropic radiation \cite{Andersson2015}.
If grain alignment is driven by an incident radiation field, its effectiveness should decrease with increasing extinction \cite{Lazarian2007,Andersson2015}.  A systematic decrease in polarization fraction towards high-extinction lines of sight (often called a `polarization hole') is indeed commonly observed  \cite[e.g.][]{Kwon2018,Soam2018,Coude2019}, implying loss of grain alignment, an effect most pronounced in starless cores which have no internal source of photons \cite{Jones2015, Jones2016}.
However, non-Gaussian noise properties of polarization measurements can lead to a predisposition for a lack of grain alignment to be inferred at low-to-intermediate signal-to-noise \cite{Pattle2019}.
Higher-sensitivity observations will allow observation of starless and protostellar cores in a wider range of environments, elucidating the conditions under which grains lose alignment with the magnetic field, and so constraining the size distribution of grains in high-density regions.

Simple models predict an approximately flat submillimetre polarization spectrum (the variation of polarization fraction with wavelength) in molecular clouds \cite{Hildebrand1999}. However, polarization spectra have been found to have a minimum at 350~$\mu$m \cite[e.g.][]{Hildebrand1999,Vaillancourt2008}.
While this is possible in the RAT paradigm \cite{Bethell2007}, the predicted variation in polarization fraction with wavelength is too small to explain the apparent 350\,\mum\ minimum.
However, recent work combining BLAST-Pol 250--500~$\mu$m and Planck 850~$\mu$m observations have shown a polarization spectrum which is flat to within 10--20\% across the submillimetre in nearby molecular clouds \cite{Gandilo2016, Shariff2019}.  Comparing polarization spectra observed on $\sim 5$\arcmin\ scales with those observed on $\sim 10^{\prime\prime}$ scales will allow investigation of how grain properties vary from molecular cloud scales to filament/core scales, and of the dependence of grain growth on temperature, density and radiation field.

\textbf{450\mum\ science:} Replacing Planck's 5\arcmin\ beam with the JCMT's 12\arcsec\ resolution will produce detailed submillimetre polarization spectra through synthesis with BLAST-TNG~\citep{BLASTTNG} and SOFIA~\citep{Harper2018}, with 36\arcsec-resolution (BLAST 500~$\mu$m). The upgrade to include 450~$\mu$m imaging could replace BLAST-TNG's 500~$\mu$m band, further improving the resolution.

\section{Synergy with molecular line magnetic field observations}
\label{sec:850methods_lines}

\textbf{Zeeman Effect:}
The most direct measurement of astrophysical magnetic field strength is through Zeeman splitting of paramagnetic spectral lines. Such line-of-sight field strength measurements, using either thermal lines (e.g., HI, OH and CN) or maser lines (e.g., H$_2$O), have only been achieved toward a limited number of sources
\cite[e.g.][]{Crutcher2012}.
Although magnetic field strengths inferred from dust emission using DCF analysis are comparatively indirect, they can, through wide-area mapping, simultaneously provide both the field structure and its strength in the plane of the sky \cite[e.g.][]{PattleFissel2019}.
While Zeeman and DCF measurements probe different magnetic field components, the two are broadly consistent with and complement each other (Figure~\ref{fig:zeeman_vs_cf}) \cite{PattleFissel2019}, and can be combined to estimate total magnetic field strength \cite{Kirk2006}. A more complex approach is to combine polarization, Zeeman and ion-to-neutral molecular line width ratios in order to determine the angle of the magnetic field with respect to the line of sight \cite{Houde2002,Houde2011}.

With its enhanced sensitivity, the new JCMT camera will allow estimates of magnetic field strengths to be obtained even in the lower-density periphery of molecular clouds. We will therefore have a more complete knowledge of magnetic field strengths over a larger range of gas densities, as shown in Figure~\ref{fig:zeeman_vs_cf}.  Polarization maps made using the new camera will thus provide a valuable reference for the planning of future measurements of the Zeeman effect as well as providing opportunities to infer three-dimensional magnetic field properties, and driving associated theoretical studies.

\textbf{Goldreich-Kylafis effect:}
Molecular line polarization can arise from the Goldreich-Kylafis (GK) effect \cite{Goldreich1981}, in which molecular line emission may in certain circumstances be linearly polarized either parallel or perpendicular to the plane-of-sky magnetic field.
The GK effect can complement emission polarimetry, occurring in regions where polarized dust emission is to faint to detect: for example, dust polarization can be used to probe magnetic fields in the high-column-density circumstellar material around YSOs, while fields in the low-column-density lobes of molecular outflows can be probed using the GK effect.
Spectropolarimetric observations can also be used to probe the structure of magnetic fields in position-position-velocity space in order to, for example, disentangle the magnetic field of the galactic spiral arms.
Wide-field Galactic plane observations made using the new JCMT camera, complemented with targeted observations of the GK effect with the forthcoming heterodyne array, might for the first time reveal the magnetic field structures of individual spiral arms, and the connection between galactic-scale and cloud-scale magnetic fields.

\section{Synergies with other polarimeters}
\label{sec:instruments}

As discussed in Section~\ref{sec:intro}, and elsewhere above, the resolution, field of view and mapping speed of the JCMT makes it an excellent bridge between $Planck$ all-sky polarization maps and interferometric polarization imaging with ALMA and other similar instruments such as the Submillimeter Array (SMA).  The new JCMT camera will both perform targeted high-resolution follow-up of $Planck$ observations and undertake wide-area surveys from which targets for interferometric follow-up can be selected.  The new JCMT camera will also perform pathfinder science for forthcoming large radio telescopes such as the Square Kilometer Array (SKA) and the next generation VLA (ngVLA).

The new JCMT camera will also synergize with other current and forthcoming single-dish polarization instruments \cite{PattleFissel2019}.  HAWC+ \cite{Harper2018}, currently operating on the airborne SOFIA observatory, is an FIR polarimeter operating in the wavelength range 53--214\mum\ with resolution 4.8--18.2\arcsec.  HAWC+ is optimized to observe a warmer dust population than the JCMT; as discussed in Section~\ref{sec:850methods_dust}, synthesis of polarization observations across the FIR/submillimetre regime is essential to understanding how  dust properties and magnetic fields vary with gas temperature and density.  BLAST-TNG \cite{Galitzki2014}, a balloon-borne polarimeter planned to fly from Antarctica in December 2019, operates at 250--500\mum\ with resolutions 30--60\arcsec.  Flight in Antarctica mean that BLAST-TNG can observe a limited number of Southern-sky targets, with declinations largely not observable by the JCMT, making the two instruments complementary.

TolTEC \cite{Bryan2018}, the camera currently being commissioned at the LMT, will operate at wavelengths 1.1--2.1mm, with resolution 5.0--9.8\arcsec.  NIKA-2 \cite{Adam2018}, a camera currently being commissioned at the IRAM 30m telescope, will operate at 1.2 and 2.0mm with resolutions of 11 and 18\arcsec.  Both cameras will offer a polarization mode.  Synthesis of JCMT 850\mum\ and 450\mum\ observations with these data will further add to polarization spectra across the wavelength range of dust continuum emission, while the higher surface brightness of cold dust at 450\mum\ and 850\mum\ will enhance the JCMT's ability to detect cold and dense sources over that of millimetre cameras.  Moreover, comparison of millimetre and submillimetre observations of sources with significant non-thermal emission will allow the effects of synchrotron radiation to be disentangled from continuum emission (c.f. Section~\ref{sec:850sci_galaxies}).  The A-MKID camera \cite{Otal2014}, currently being commissioned at APEX, will operate at 350\mum\ and 850\mum\ with resolutions of 8 and 19\arcsec, and will offer a polarization mode.  Observing declinations $<+50^{\circ}$, A-MKID may prove to be an effective Southern-sky counterpart to the JCMT.

All of these single-dish polarimeters modulate signal using a half-wave plate or polarizing grid \cite{PattleFissel2019}.  The new JCMT camera will therefore have an intrinsic advantage in its ability to measure polarization natively, making it unique in its ability to provide polarization information as a standard component of an astrophysical observation.

\section{Executive Summary} 
\label{sec:summary}

The James Clerk Maxwell Telescope (JCMT), which has long been at the forefront of submillimetre polarization instrumentation, is proposing a next-generation 850 \mum\ camera.  In this white paper we have presented the science case for the polarimetric capabilities of this camera.
The JCMT's current POL-2/SCUBA-2 system has provided its user community with an outstanding and unique imaging polarimeter, and has resulted in numerous international collaborations and the global exchange of knowledge and ideas.  The science goals and instrumentation requirements described in this work are based on wide-ranging discussions both across and beyond the JCMT community.

The new JCMT camera will be a vital tool with which to address fundamental questions of the role of magnetic fields in galactic astronomy and star formation,
and also to pursue such studies beyond the Milky Way by analysing the magnetic field properties of galaxies as a whole. Key questions in such galactic and extra-galactic studies would include that of the role of magnetic fields in determining star formation efficiency and the origin of the Initial Mass Function, and in determining a galaxy's structure, ISM thermal balance, and global star formation rate.
The JCMT is vital to such studies, with a mapping size of tens of arcminutes and an angular resolution of 14\arcsec at 850\mum, and operating in the optimal wavelength regime for detection of cold and dense material. 
The new JCMT camera's enhanced sensitivity and wide field of view will
make it unique in its ability to serve as a wide-area survey instrument with which to study the interplay between self-gravity, turbulence and magnetism which drives the evolution of the cold ISM, and so to resolve questions crucial to our understanding of cosmic star-formation history.

\bibliographystyle{unsrtMaxAuth}

\end{document}